\documentclass[aps,prb,amsmath,amssymb,footinbib,showpacs,twocolumn]{revtex4-2}
\usepackage{amsmath}
\usepackage{amssymb}
\usepackage{amsthm}
\usepackage{setspace}
\usepackage{graphicx}
\usepackage{braket}
\usepackage{mathrsfs}
\usepackage{float}
\usepackage[colorlinks = true,linkcolor = blue,urlcolor  = blue,citecolor = blue,anchorcolor = blue]{hyperref}
\usepackage[utf8]{inputenc}
\usepackage[english]{babel}
\usepackage{bm}

\begin{document}


\title{Magnetoresistance in Magnetic Weyl semimetal Mn$_{3}$ZnC}
\author{Sunil Gangwar}
\author{C. S. Yadav}
 \email{shekhar@iitmandi.ac.in}
\affiliation{School of Physical Sciences, Indian Institute of Technology Mandi, Kamand, Mandi-175075 (H.P.) India$^1$}%
\affiliation{Center for Quantum Science and Technologies, Indian Institute of Technology Mandi, Kamand, Mandi-175075 (H.P.) India$^1$}%
\date{\today}

\begin{abstract} 

The magnetoresistance (MR) in magnetic materials reveal an intriguing spin-dependent electron scattering, highlighting the role of spin polarization on the electronic transport properties. This becomes even more interesting for the topological magnetic materials where a finite Berry curvature leads to an intrinsic scattering channel also. In this report, we investigate MR of Mn$_{3}$ZnC, an antiperovskite magnetic nodal line semimetal. Mn$_{3}$ZnC shows a ferromagnetic (FM) transition at $\sim$ 420 K, followed by an ferrimagnetic (FIM) transition at $\sim$ 195 K. We focus on the interpretation of MR data and highlight the relation between MR and its magnetization. The signature of magnetization induced MR shows correlation at low magnetic fields and displays distinct behaviors in the FIM and FM states. Similar to the isothermal magnetization M(H) curve of FM and FIM state, the MR curves exhibit a sharp increase, followed by a linear behavior at higher fields. As the magnetic field varies, the cusp-like anomaly becomes more pronounced, and vanishes at the magnetic phase transition, which then reappears as the temperature increases. The sign change in the MR curves in the FIM state is attributed to a drastic change in the carrier mobility. Interestingly, this system shows a positive MR in FM state at very low field, which is quite unusual.

\end{abstract}

\maketitle

\section{Introduction}
The study of magnetoresistance (MR) in metallic magnets is vital area of research fueled by both, the fundamental scientific interest and wide-ranging technological applications. \cite{zhou2020linear,daughton1999gmr}. For the past several decades, many new materials have been discovered that exhibit large MR; which is the relative change in electrical resistivity of a material in the presence of an applied magnetic field \cite{zhou2020linear,bombor2013half,gerber2007linear}. MR is defined as  $ MR \%= [(\rho (H) - \rho (0))/\rho (0))] \times 100$, where $\rho(H)$ denotes the resistivity measured in the presence of a magnetic field, and $\rho(0)$ represents the resistivity in the absence of a magnetic field. Materials have fixed resistivity at a given temperature, but magnetic field influences the motion of charge carriers, and resistivity value changes (increase/decrease). 

Generally, non-magnetic materials exhibits positive MR, which is primarily attributed to the Lorentz force acting on charge carriers, and MR shows a quadratic dependence on field (MR $\sim$ H$^{2}$) at low magnetic field \cite{gerber2007linear,zhou2020linear,yamada2019experimental}. In charge compensated systems, the quadratic behavior of MR extends up to high-field regime. Whereas, in non-compensated systems, MR typically saturates at high fields, as the dominant charge carriers no longer contribute to a further increase in resistivity \cite{murgatroyd2021experimental}. For magnetic systems, field dependence of MR varies depending on the orientation of spins responsible for the scattering with conduction electrons. However, the MR in magnetic topological materials systems is quite complex, primarily due to the varying distribution of grain and domain sizes, complexity of the electronic band structure \cite{nath1998magnetoresistance,gomonay2002magnetostriction}. In ferromagnetic (FM) systems, external magnetic field strongly aligns the magnetic domains, reduces domain walls and the associated \textit{s-d} scattering, resulting in the lower resistivity and hence a negative MR \cite{gomonay2002magnetostriction,yamada2019experimental}. In antiferromagnetic (AFM) materials, magnetic field disrupts the spin alignment and enhances the spin-flop transition, resulting in an increase in resistivity that leads to a positive MR \cite{nath1998magnetoresistance}.  The magnetization of material also plays an role in influencing the spin polarization of electrons and their scattering behavior. In 1997, Xu \textit{et al.} showed a linear magnetoresistance (LMR) in doped silver chalcogenides (Ag$_{2+\delta}$Te and Ag$_{2+\delta}$Se; $\delta$ = 0.1) starting from 10 Oe to 55 kOe magnetic field in the wide temperature range of 5 - 300 K \cite{xu1997large}. The observation of LMR in these system led to the development of various theories for such behavior. Later many other correlated systems like graphene \cite{wang2014classical}, and Heusler alloys \cite{yamada2019experimental} were found to exhibit a non-saturating LMR even at low magnetic field. The LMR has been understood through various mechanisms such as open orbit in Fermi surface, mobility fluctuations (Disorder model of Parish-Littlewood), and extreme quantum limit (Abrikosov's quantum theory, where only the lowest Landau level is occupied) etc. \cite{laha2020magnetotransport, huang2016linear}.

In this study, we have explored the MR in Mn$_{3}$ZnC, a carbide Mn-based antiperovskite and a topological nodal-line semimetal that exhibits both FM and ferrimagnetic (FIM) phases \cite{teicher2019weyl}. The band structure calculation on the compound suggested it as magnetic Weyl semimetal with 24 Weyl nodal points in the unit cell. Both the long range magnetic ordering and topological feature make this an interest system for MR studies. The polycrystalline Mn$_{3}$ZnC used in this study exhibits a paramagnetic to FM transition at $\sim$ 420 K, followed by a ferrimagnetic (FIM) transition at $\sim$ 195 K \cite{gangwar2024magneto,gangwar2025berry}. In our previous study, we showed that finite Berry curvature is the primary driving mechanism behind the anomalous Hall and Nernst effects in Mn$_{3}$ZnC \cite{gangwar2025berry}. Here, we present a comprehensive study of the MR in Mn$_{3}$ZnC, addressing the high-field linear LMR in the framework of both classical and quantum theories, and further examine the influence of magnetic states on the MR at low applied magnetic fields. 

\section{Results and discussion}

Figure 1 (a, b) show the transverse MR, when the applied magnetic field is perpendicular to current at various temperatures, lower and higher than FIM transition respectively. We examine the MR data in the high and low magnetic fields regimes across the FIM transition, as follows;

\begin{figure*}
\includegraphics[width= 14 cm, height = 12 cm]{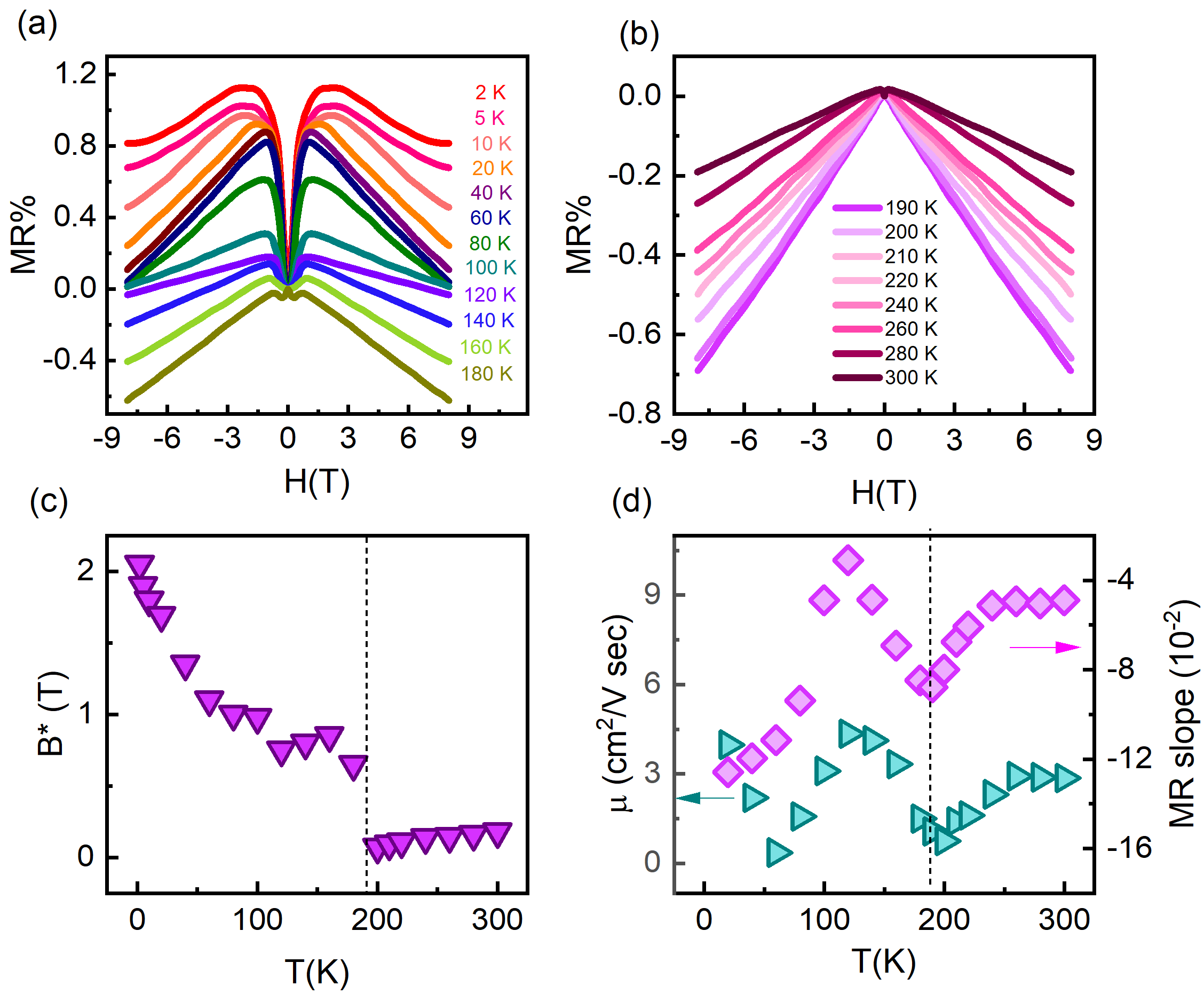}
\caption{Magnetic field dependent magnetoresistance
(MR) of Mn$_{3}$ZnC in the temperature range (a)  1.8 - 180 K, and (b) 190 - 300 K. (c) Critical magnetic field $B^*$ as a function of temperature. (d) Carrier mobility $\mu$ and MR slope as a function of temperature.}
\label{fig:Figure1}
\end{figure*}

\subsection{High-Field Magnetoresistance}

Mn$_{3}$ZnC shows negative MR for H $\ge$ 2 T, for 1.8 $\le$ T $\le$ 300 K. In the temperature range $T$ = 1.8 to 5 K, the MR curves exhibit a saturating trend above H = 7 T, and an increase in magnetic field have negligible effect on the $s-d$ scattering. The magnetization in Mn$_{3}$ZnC saturates at a field of 0.2 - 0.3 T in this temperature range \cite{gangwar2025berry}, indicating the competing mechanisms for MR. Above 10 K, the MR curves become linear, which is an interesting characteristics. Generally, the linear MR (\textit{MR $\propto$ H}) in materials is understood through the mechanisms such as: 
(i) Parish-Littlewood (classical disorder) model, where the mobility fluctuations of charge carriers are considered to be responsible for LMR. Distortion in the carrier path induced by disorder led to the mixing of Hall component to longitudinal resistance \cite{parish2003non}. The multiple scattering of high mobility carriers generate a drift velocity by low mobility islands perpendicular to the electronic cycloidal trajectories in an interaction of applied electric and magnetic field. This model proposes two possible conditions; one of which is the crossover magnetic field $B^*$, at which the MR begins to exhibit linear behavior, should be inversely proportional to the carrier mobility $\mu$ ($B^*$ $\propto$ 1/$\mu$). Second, the MR slope should be proportional to the carrier mobility (MR slope $\propto$ $\mu$) \cite{parish2003non,huang2016linear,feng2015large}. The temperature variation of $B^*$ (Fig. 1(c)), exhibits a sharp decrease as the temperature rises, followed by an almost temperature-independent behavior. A noticeable change in slope occurs near the magnetic transition temperature 195 K. The carrier mobility $\mu$ is also shown in Fig. 1(d) as a function of temperature, exhibiting a trend opposite to that of $B^*$ with respect to mobility. The slope of MR, shown in Fig. 1(d), exhibits the same temperature dependence as the mobility. Both these conditions as outlined in the PL model are satisfied in this case. However, our sample forms in a single homogeneous phase and but exhibits low mobility. Therefore, the observed non-saturating linear MR cannot be attributed to the classical model. Considering the fact that the studied compound has low mobility (1 - 4 cm$^{2}/$V-sec), with magnetic and topological nature, the applicability of this model is high unlikely.

\begin{figure*}
\includegraphics[width= 14 cm, height = 12 cm]{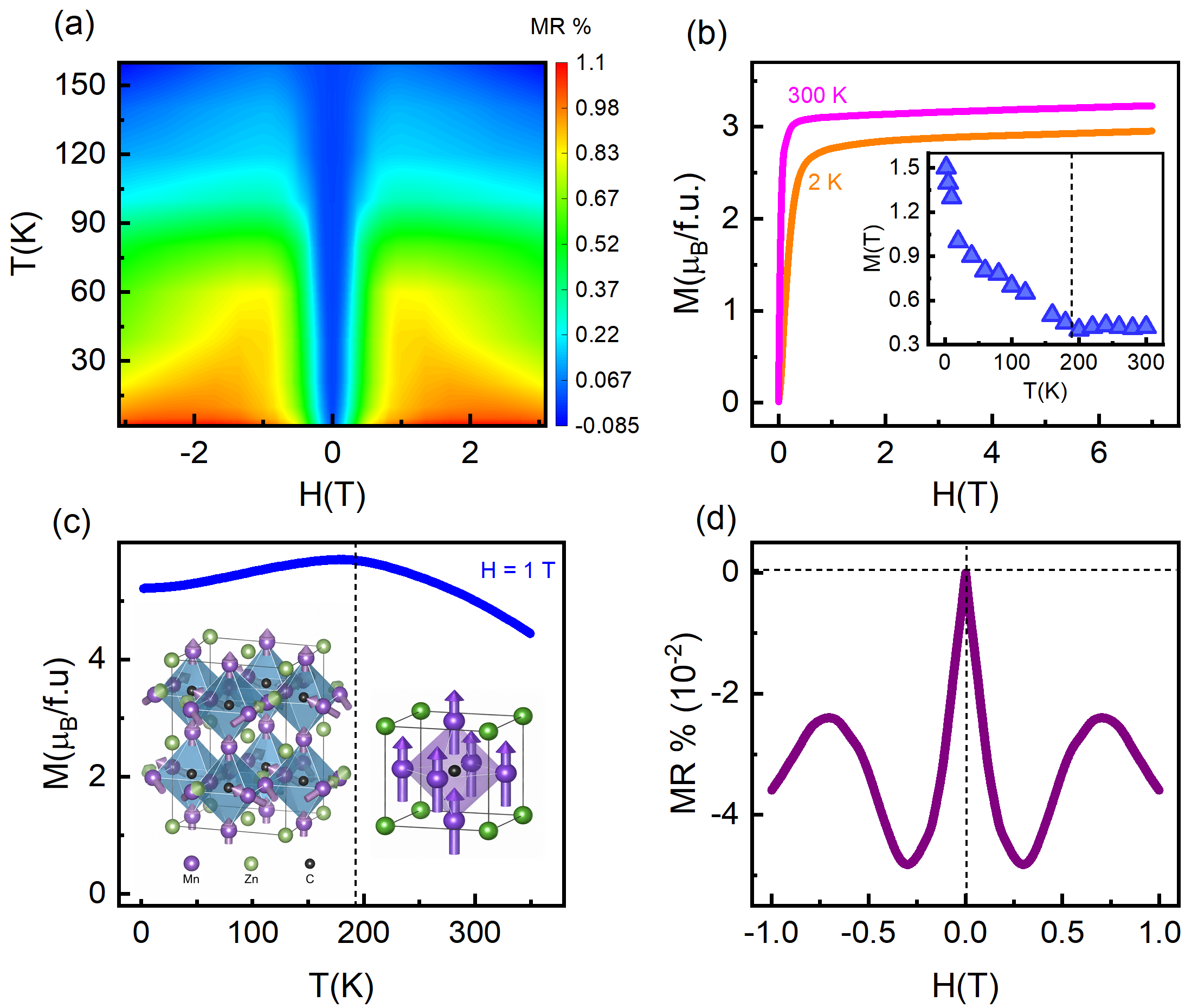}
\caption{(a) Contour plot of MR \% in the temperature range of 1.8 - 160 K up to 3 T applied magnetic field. (b) Isothermal magnetization as a function of magnetic field at 2 K and 300 K. Inset shows the critical magnetization values as a function of temperature. (c) Magnetization as a function of temperature at H= 1 T applied magnetic field. Inset shows the magnetic crsytal structures in ferrimagnetic and ferromagnetic states. (d) MR \% as a function of temperature at 180 K.}
\label{fig:Figure2}
\end{figure*}

(ii) Guiding Center Diffusion Model, where in low disorder topological material, the guiding center of the cyclotron motion of electron can also move randomly, and the diffusion of these centers lead to LMR.  However, this mechanism is also not applicable owing to low mobility of the system \cite{song2015linear,li2026quantum}.

(iii)  Open Orbits in Fermi surface, may also lead to LMR in the intermediate magnetic field range. Though the intermetallics like Mn$_{3}$ZnC have connected Fermi surface, truncated at Brillouin zone boundary, observation of this effect in a polycrystalline material is also unlikely \cite{silvera2026fermi} 

(iv) Abrikosov’s Quantum theory of LMR, accounts for the LMR in the system with gapless, linear Energy dispersion (E $\propto$ k), and mostly considered for topological materials such as Dirac, Weyl semimetals. Abrikosov showed that in the extreme quantum limit when the electrons are confined to lowest Landau level and the cyclotron energy is larger than the Fermi energy of system and thermal energy, MR increases linearly with field \cite{abrikosov1998quantum, huang2016linear, abrikosov2003quantum}. In this regime, the density of states (DOS) is highly quantized, with each Landau level contributing a small portion of the total DOS. As the magnetic field increases, DOS also rises, which in turn leads to an increase in the current and leads to LMR in a system. Abrikosov calculated the electrical resistivity for an isotropic metal in the extreme quantum limit, when only the lowest Landau level is filled, as \cite{abrikosov2003quantum, abrikosov1998quantum}
\begin{equation}
 \rho_{xx} = (N_i H/ \pi n^2 ec)  
\end{equation}

where $\rho_{xx}$ is resistivity, \textit{n} is carrier concentration, N$_{i}$ is the concentration of scattering centers and $c$ is the speed of light. The condition for this expression is as \cite{abrikosov2003quantum,huang2016linear,abrikosov1998quantum}
\begin{equation}
 n<< (eH/\hbar c)^{3/2}  
\end{equation}
where $e$ and $\hbar$ represent the electric charge and Planck's constant, respectively. For H $\sim$ 10 T, the carrier concentration corresponds $\sim 10^{18} \text{cm}^{-3}$. This value is lower than $ \sim 10^{22} \text{cm}^{-3}$ observed for Mn$_{3}$ZnC \cite{gangwar2025berry}. However, owing to the multiband nature and weakness of the pesudopotential for electron-impurity interaction, this condition can be fulfilled at accessible lower fields also \cite{abrikosov2000quantum}.

Recently, LMR is reported in the other topological semimetal Cd$_3$As$_2$ \cite{laha2020magnetotransport} CaCdSn \cite{liang2015ultrahigh} in the similar field regime, and we believe that the non-trivial band structure (linear energy dispersion) in these compounds including that of Mn$_3$ZnC may be responsible for the LMR, and may be accounted through Abrikosov' quantum limit theory.

\subsection{Low-field Magnetoresistance}

In this section, we examine the correlation between low-field MR and magnetization in FM and FIM states. Figure 2(a) shows a contour plot of MR in the temperature range of 1.8 - 160 K up to 3 T field. MR increases rapidly at low-field and up to a field ($B^*$), then begins to decrease as the field continues to rise. This behavior is similar to isothermal magnetization data, where $M$ increases rapidly before saturating at higher field (Fig. 2(b)). The critical field $B^*$ where the MR starts to drop sharply, closely corresponds to the critical magnetization ($M^*$) as shown in the inset of Fig. 2(b), where $M^*$ is estimated from the linear extrapolation of the high field M(H) curves. When the field is strong enough to cause saturation, all spins align in parallel, leading to a reduction in $s-d$ scattering and decrease in resistivity, as reflected through $B^*$. Similar results have been observed in other magnetic materials as well \cite{shinjo1990large}. Figure 2(c) shows the temperature-dependent magnetization measured under at magnetic field of H = 1 T. The system exhibits a cubic FM state and a noncollinear FIM tetragonal phase, as shown in the magnetic structure illustrated in the inset. The magnetic moments in a AFM or FIM system, suppresses as the field increases, while the FM order stabilizes at higher fields. At the fields, above $B^*$, magnetization reaches saturation state, corresponding to an ordered state. As a result, $s-d$ scattering is reduced, causing a decrease in MR. Near FIM transition ($\sim$ 180 K), we observed an abrupt change in (negative) MR  for $H$ $\le$ 0.8 T leading to a sharp V-shaped peak (Fig. 2(d)). This response is plausibly associated with magnetization related effects such as domain wall motion or spin-disorder suppression near a magnetic phase boundary \cite{gomonay2002magnetostriction}.

\begin{figure*}
\includegraphics[width= 14 cm, height = 12 cm]{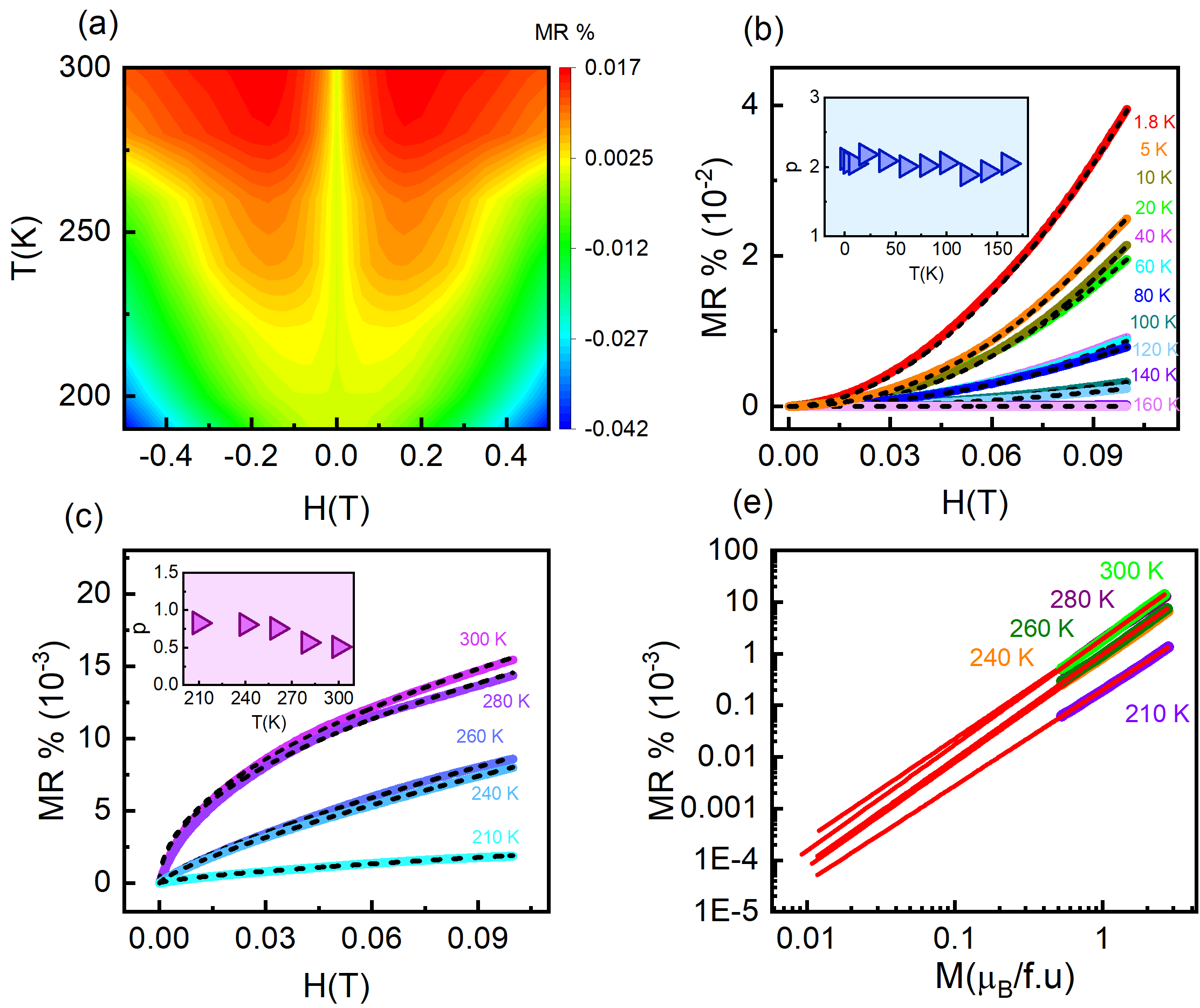}
\caption{(a) Contour plot of MR \% in the temperature range of 190 - 300 K up to 0.5 T applied magnetic field. (b) MR \% curves as a function of field in the temperature range of $T$ = 1.8 – 160 K, fitted with MR $\propto$ $H^{p}$ power law. (c) MR \% curves as a function of field in the temperature range of $T$ = 210 – 300 K, fitted with MR $\propto$ $H^{p}$ power law. The insets in (b) and (c) show the temperature dependence of the exponent $p$. (d) MR \% curves as a function of magnetization in the temperature range of T = 210 – 300 K, fitted with MR $\propto$ $M^{\beta}$ power law.}
\label{fig:Figure3}
\end{figure*}

We have shown a contour plot of MR at low-field in FM state in the temperature range of 190 - 300 K and magnetic fields up to 0.5 T, in Fig. 3(a). The observed negative MR in the FM state can be attributed to the spin-dependent scattering and the suppression of spin disorder scattering. In the absence of an applied field, randomly oriented magnetic domains lead to electron scattering and consequently, the resistivity. When a magnetic field is applied, these domains align with the field, reducing spin disorder. This alignment decreases scattering, reduces resistivity, and results in negative MR \cite{bombor2013half,nath1998magnetoresistance}. At 190 K, the MR exhibits an almost linear behavior across the entire applied field. Above 190 K, the MR curves show a weak cusp-like behavior with positive values at low magnetic fields, and these positive values increase with rising temperature. However, this cusp-like behavior diminishes as the system transitions from FIM state to FM state at $\sim$ 190 K. As the temperature approaches room temperature, a cusp-like behavior begins to increase once again. The development of cusp-like behavior with positive MR in the FM at very low-field is indeed unusual, it is most likely arising from the spin-dependent scattering of electrons with phonons, magnons, defects and impurities \cite{ahmed2019magnetic,bombor2013half}. Bombor $et$ $al.$ proposed that, at high temperatures, magnons are the primary scattering centers responsible for the MR in Co$_{2}$FeSi \cite{bombor2013half}. They suggested that the magnetic field reduces the number of electrons available for scattering while simultaneously increasing the energy of the magnon. At low magnetic fields, electron-electron scattering, electron-phonon scattering, and electron-defect scattering become significant, contributing to the positive MR.

we have fitted the low-field MR data in the different magnetic states using separate scaling models. Specifically, we analyzed the field dependence of MR using the relation MR $\propto$ $H^{p}$, where the exponent $p$ provides insight into the dominant magnetic correlations and spin-scattering mechanisms \cite{nath1998magnetoresistance}. In non-magnetic systems, compensated materials typically exhibit $p$ = 2, whereas uncompensated systems yield $p$ = 1. In magnetic materials, however, $p$ is not universal; it depends strongly on temperature, magnetic ordering, and the underlying scattering processes. In the FIM state, we obtain $p$ $\sim$ 2, consistent with a Lorentz-type MR, as shown in Fig. 3(b). We note that other FIM systems, such as Sr$_{2}$CrMoO$_{6}$ thin films have reported $p$ $\sim$ 1, attributed to spin-dependent scattering \cite{wang2021giant}. This highlights that in magnetic systems the exponent $p$ can vary depending on the specific scattering mechanism. 
In the FM state, the MR follows an $p$ $<$ 1 dependence on magnetic field, as shown in Fig. 3(c). It indicates that MR is not governed solely by Lorentz-force-induced orbital motion. Instead, the dominant contribution arises from spin-dependent scattering and magnetic-ordering effects. As external magnetic field is applied, spins progressively align with the field, which lead to the reduction in the spin scattering. The reduction in scattering becomes progressively weaker with field, giving rise to a sublinear field dependence. Similar sublinear exponents ($p$ $<$ 1) have been reported in FM systems, including Fe$_{80-x}$Ni$_{x}$Cr${_20}$ alloys \cite{nath1998magnetoresistance}, AuFe \cite{nigam1986magnetoresistance}, and NiMn \cite{senoussi1984anomalous}.

To examine the correlation between magnetization and MR under isothermal conditions, we used the empirical relations MR $\propto$ $H^{p}$ and M (H) $\propto$ $H^{q}$. These relations yield MR $\propto$ $M^{\beta}$, where the exponent $\beta$ = p/q has frequently been found to be close to 2 for FM systems \cite{nath1998magnetoresistance,nigam1986magnetoresistance}. Consistent with these observations, Mn$_{3}$ZnC also exhibits an $\beta$ value of $\sim$ 2, and our analysis follows this established scaling behavior, as shown in Fig. 3(d). To ensure reliability of the extracted exponents, all fits were performed within a low-field range up to H = 0.1 T.

Figure 4 shows the temperature dependence of MR, which decreases monotonically as the temperature increases. The value of MR is $\sim$ 0.8 \%, at 1.8 K, and reaches to - 0.23 \% at room temperature. The magnitude of MR is small, owing to the low carrier mobility in the compound. A sign reversal is observed around 120 K, which coincides with a sharp drop in carrier mobility at roughly the same temperature. The sign change in MR observed in various materials and has been associated with the interplay between carrier mobility and scattering mechanisms \cite{li2020mobility,hu2008classical}. Similarly, for Mn$_{3}$ZnC, the carrier mobility changes abruptly around 120 K, as shown in Fig. 1(d), which may be responsible for the sign change in the MR. A distinct drop is observed around the magnetic phase transition ($\sim$ 195 K). Above this temperature, the MR begins to decrease with increasing temperature, as expected for the FM state.

\begin{figure}
	\begin{center}
		\includegraphics[width=7.0cm]{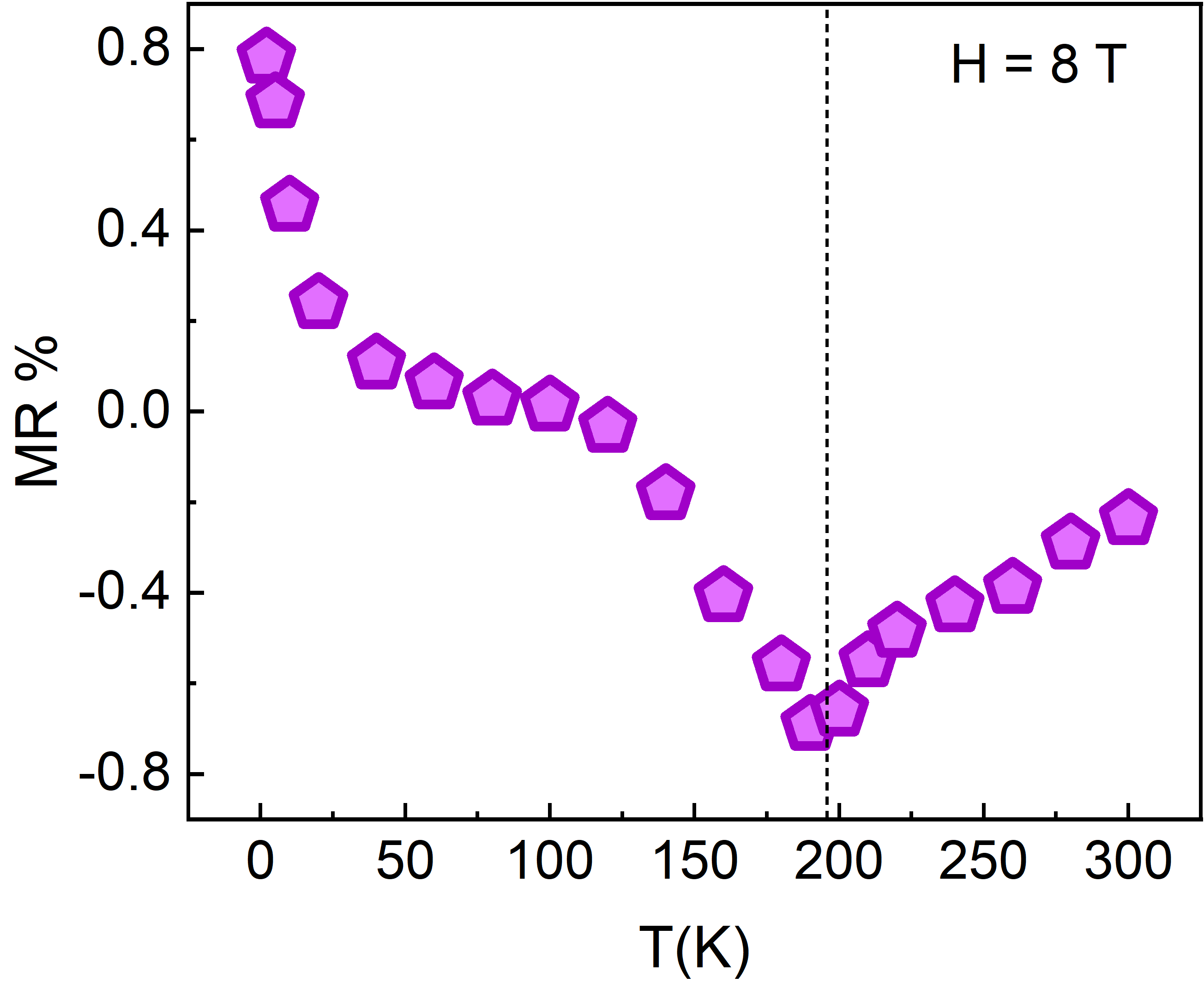}
		\caption{\label{Fig4}Magnetoresistance as a function of temperature at H = 8 T field.}
	\end{center}
\end{figure}

\section{conclusion}
In summary, we performed an analysis of transverse MR in magnetic Weyl semimetal Mn$_{3}$ZnC. We observed magnetization-induced MR, where the MR curves closely resemble the magnetization curves at low applied magnetic fields. This behavior is indicative of the coupling between the magnetic ordering and the electronic transport in the system. Owing to spin-dependent scattering, the MR exhibits distinct magnetic-field dependencies: a quadratic dependence in the FIM state and a sublinear dependence in the FM state. The change in the sign of the MR at 120 K is associated with the abrupt variation in carrier mobility observed near the same temperature. In the high-field region, the MR curves show a non-saturating linear behavior, which is understood through Abrikosov's quantum limit theory. The linear energy dispersion and the pseudopotential associated with electron-impurity interactions may be responsible for the applicability of the quantum theory of linear LMR in this system. These results highlight the complex magnetic interactions and their influence on the electronic transport in the system.

\section{Acknowledgment}
We acknowledge Advanced Material Research Center (AMRC), IIT Mandi for the experimental facilities. SG and CSY acknowledge IIT Mandi and India for the HTRA fellowship. This research received no external funding.

\section{Author contributions}
SG prepared the material and performed the experimental measurements and wrote the manuscript. CSY supervised the overall project.

\bibliography{Mn3ZnC}

\end{document}